\documentclass[prb,twocolumn,floatfix,showpacs]{revtex4}
\usepackage{graphicx,natbib}
\begin{document}


\title{Low Temperature Static and Dynamic Behaviour of the Easy-Axis
Heisenberg Antiferromagnet on the Kagome Lattice}


\author{S. Bekhechi and B.W. Southern}
\affiliation{Department of Physics and Astronomy\\ University of Manitoba \\ Winnipeg Manitoba \\ Canada R3T 2N2}


\date{\today}

\begin{abstract}
The antiferromagnetic Heisenberg model with easy-axis exchange anisotropy on the Kagome
lattice is studied by means of Monte Carlo simulations. From equilibrium
properties, we find that the values of the critical exponents associated with the
magnetization at the critical temperature $T_c$ vary with the magnitude of the anisotropy. On the other hand, 
the spin-spin autocorrelation functions have a stretched
exponential behavior with a power law divergence of the relaxation time at a glass-like temperature $T_g \sim T_c$. From 
non-equilibrium dynamics at a fixed temperature below $T_g$, aging effects are 
found which obey the same scaling laws as in spin glasses and polymers.
\end{abstract}


\maketitle

\section{Introduction}

The Kagome spin system has attracted much interest, both theoretically and
experimentally. Because the geometry of the lattice consists of corner sharing triangles in a
layer which surround hexagons,  spin systems are
highly frustrated when antiferromagnetic interactions are present \cite{n1}. In
the case of the nearest-neighbour antiferromagnetic spin $1/2$ Ising model, the ground state is disordered and the spin-spin correlation
function decays exponentially \cite{n2a} at zero temperature. This behaviour differs from that which occurs in
 other  periodically frustrated  2d Ising lattices such as
 the triangular and  fully-frustrated square lattices where the
 correlation function decays as a power law
\cite{n2b,n2c}. Frustration also leads to a larger  macroscopic
entropy \cite{n3a}  in the Kagome lattice compared to the triangular lattice\cite{n3b}.

The classical isotropic Heisenberg antiferromagnet on the Kagome lattice also has a  
ground state with
macroscopic degeneracy.   Various perturbations such as quantum 
fluctuations\cite{n5a,n5b,n5c}, 
or the addition of other couplings including further neighbour interactions \cite{n4a}, easy plane \cite{n7} or
easy-axis anisotropy\cite{n8} and Dzyaloshinski-Moriya 
interactions\cite{n9}, have a strong effect on the ground state manifold. 
There has been  a great deal of 
controversy about whether magnetic order \cite{n4a,n4b,n4c}exists in the ground state or if it remains
disordered\cite{n5a,n5b,n5c}. Two particular coplanar and ordered states, 
called the ${\bf q}=0$ and the $(\sqrt{3}\times\sqrt{3})$
configurations, are favoured by entropy effects and this effect is often referred to as order by
disorder.  The latter ordered state\cite{n6a,n6b,n6c}  is  the one that is favoured or  perhaps even 
a mixed  disordered state of both structures \cite{n6d}. 

A chirality ${\bf \kappa}$ vector characterizes these ordered configurations and  is
defined as the pairwise vector product clockwise around a triangle:
\begin{equation}
{\bf \kappa}=\frac{2}{(3\sqrt{3})}[{\bf S_1}\times {\bf S_2}+ {\bf S_2}\times
{\bf S_3}+
{\bf S_3} \times {\bf S_1}]
\end{equation}
where $1,2,3$ label the three sites in the unit cell and form three interpenetrating triangular sublattices.
The two ordered states mentioned above  are shown in figure 1 and correspond  to
configurations with uniform (ferromagnetic)  and staggered
(antiferromagnetic) chiralities respectively. 
The  structure of these ground states
allows for the formation of collective zero-energy spin
rearrangements, called weathervane defects, that permit the system to explore the ground state manifold. These
excitations
 involve the continuous rotation of the spins on
two of the sublattices about the direction defined by the third. The defects may either
traverse the entire lattice ("open" spin folds) as in the ${\bf q}=0$ state or form localized loops
("closed" spin folds) as in the $(\sqrt{3}\times \sqrt{3})$ state \cite{n4a,n6a,n7} (see figure 1). 


  It is known that most quasi-two dimensional magnetic  materials 
exhibit some kind of spin anisotropy which may be of the  easy-axis \cite{n9a} or easy-plane\cite{n10} type.
Easy-plane 2d magnets have attracted  attention due to the possibility of  a topological
Kosterlitz-Thouless phase transition which may exhibit glassy behavior different from that found in
 conventional site-disordered systems \cite{n10}. The amount of interest devoted to easy-axis
magnetic systems has been considerably smaller, especially with regard to the
study of  dynamical properties. 
Kuroda and Miyashita\cite{n8}  (KM) have previously studied  the
Ising-like Heisenberg antiferromagnet on the Kagome lattice using Monte Carlo methods. They have shown 
the existence of a phase transition at very low temperature  with an exotic
ordered phase which has no spatial long-ranged order and hence shares some similarities to spin glasses. 
In the present work we study both the equilibrium and dynamic properties of this model.
 
The paper is organized as follows: in section II we describe the model and our methods of
calculation. Section III describes the equilibrium properties of the model and the critical exponents deduced from
finite size scaling. Results 
 for the spin-spin autocorrelation function are also given in this section. A good  fit is obtained using a functional form
 which has  been used
to describe  3d spin glasses\cite{n11} and other complex systems 
\cite{n12a,n12b,n12c}.
A characteristic timescale $\tau$
diverges with a power law at a temperature close to the static critical temperature. In section IV,
  we  study the dynamical behavior below the critical temperature. The
spin-spin autocorrelation function exhibits aging effects characteristic 
of glasses. We summarize our results in section V.

\begin{figure}
\centering
\includegraphics[height=50mm,width=80mm]{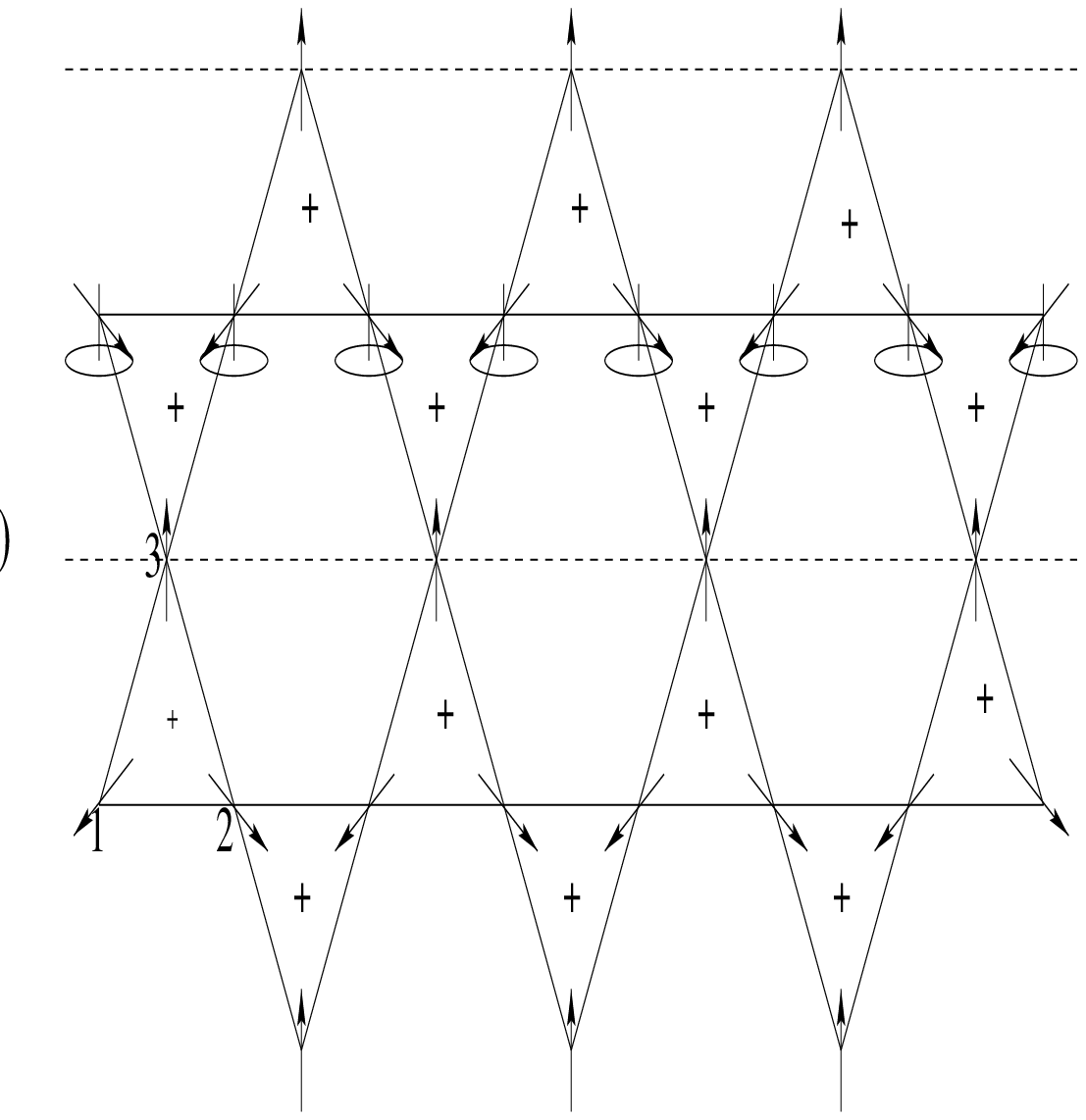}
\includegraphics[height=50mm,width=80mm]{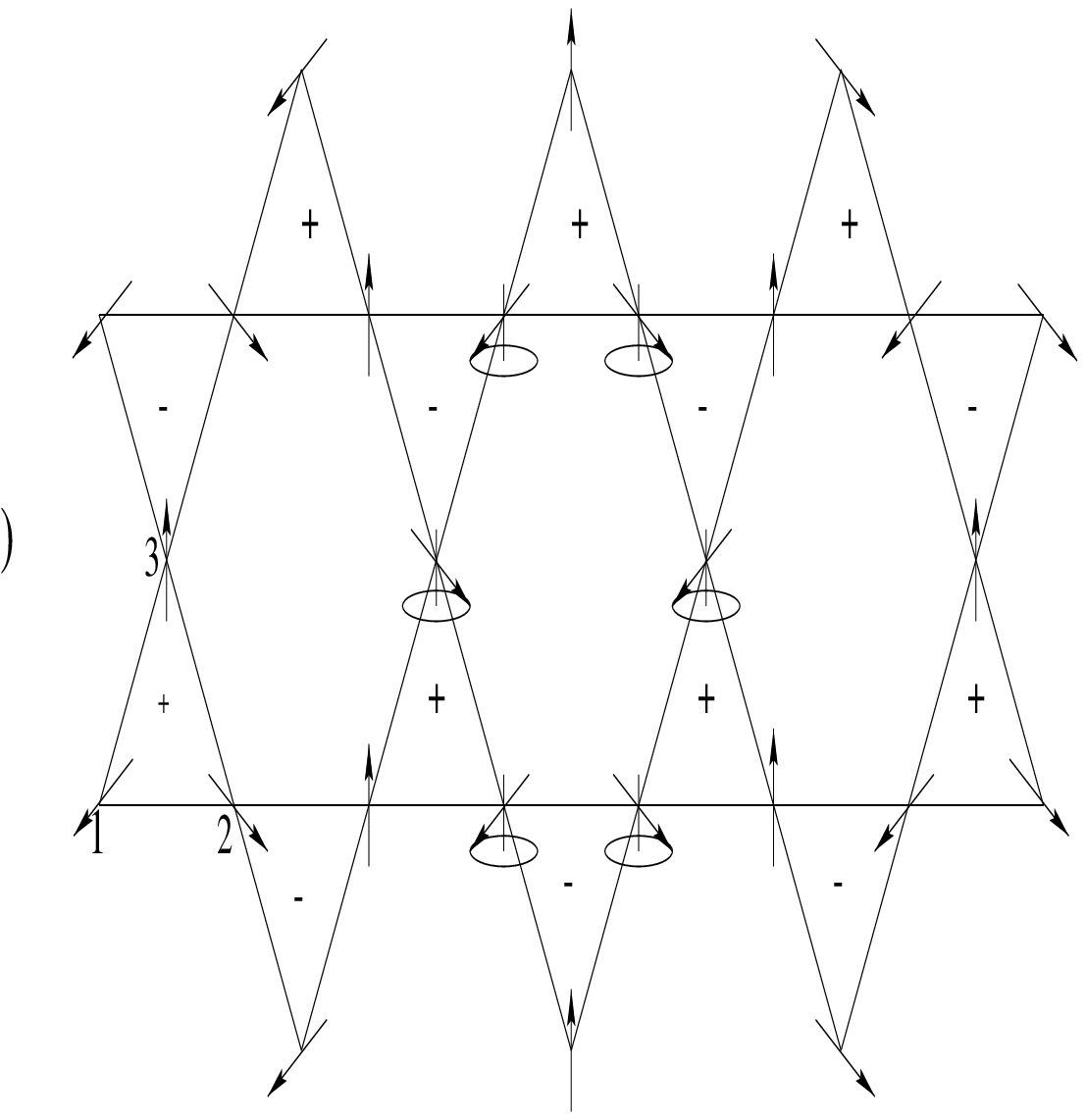}
\caption{Isotropic Heisenberg antiferromagnet on the Kagome lattice: (a) the {\bf q}=0 state in
which the spins on each of the sublattices are parallel to each other and make an
angle of $120^o$ with the spins on the other two sublattices and (b) the 
$(\sqrt{3}\times \sqrt{3})$ structure has a larger unit
cell. The $+$ and $-$ on the triangles indicate the chirality and the circles describe the open and closed spin folds in the states {\bf q}=0
and $(\sqrt{3}\times \sqrt{3})$ respectively.}
\end{figure}

\section{Model and Methods}

The model is described by the following Hamiltonian
\begin{equation}
H = J\sum_{i<j} (S_{i}^{x} S_{j}^{x} + S_{i}^{y} S_{j}^{y}+ A S_{i}^{z}
S_{j}^{z}).
\end{equation}
where $(S_{i}^{\alpha},\alpha=x,y,z)$ represents a classical three component
spin of unit magnitude located at each site $i$ of a Kagome lattice and the
exchange interactions are restricted to
nearest-neighbour pairs of sites. The parameter $A$ describes the strength of the exchange anisotropy. 
We restrict our attention to the case where  $A>1$ represents an easy-axis anisotropy. The limit $A \rightarrow 1$
corresponds to the isotropic Heisenberg model  whereas the limit $A
\rightarrow \infty$ corresponds to an infinite spin Ising model. 
 The model has a macroscopic ground state degeneracy for all $A\ge1$ with a ground state energy per site given by $-\frac{2}{3}\frac{A^2+A+1}{A+1}$. 

\begin{figure}[b]
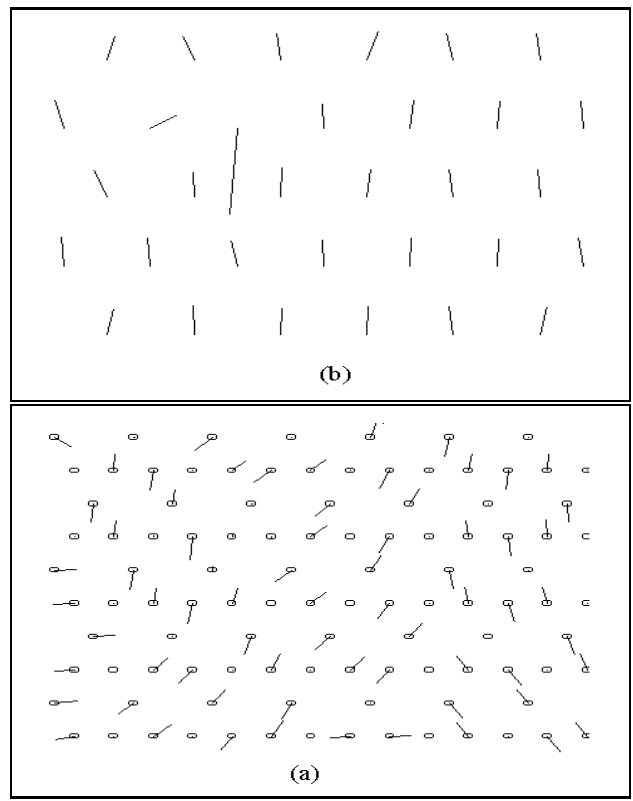

\centering
\fbox{\includegraphics[height=50mm,width=80mm]{fig2b.eps}}
\fbox{\includegraphics[height=50mm,width=80mm]{fig2a.eps}}
\caption{A Monte Carlo snapshot of the spin configurations at low $T$ for the case $A=2$. The lower panel shows the $x-y$ spin plane. On each triangle there is one spin in the z-direction but there is no spatial sublattice order. The upper panel
shows the total magnetization on each upward triangle for the same spin configuration. There is a net magnetization in the $z$-direction.}
\end{figure}


 The ground state of the system for $A>1$ corresponds to a  configuration in which the spins on each triangle form
 a distorted $120^o$ planar state  with a net nonzero magnetization in the  $\pm z$-direction whose magnitude is related to $A$ as
 $|m^z|= (\frac{A-1}{A+1})$.  The local chirality ${\bf \kappa}$ is normal to the plane of {\it each} triangle but it does not show any evidence of long range order. Miyashita and Kawamura\cite{n25} (MK) have previously studied the same model on the triangular lattice and observed a non-trivial degeneracy related to the rotation of the magnetization vector in the plane of the triangles.  In contrast to the corner-shared triangles of the Kagome lattice, the triangular lattice shares edges and there is a ${\bf q}=0$ sublattice order at $T=0$. There are also  two sequential finite temperature topological phase transitions \cite{n14b,n26} with the lower transition corresponding to the onset of a power law decay of the chirality. 
The same degeneracy arguments of MK apply to the
Kagome case but at finite temperature this degeneracy seems to be  lifted by an order from disorder effect and the $z$-axis is preferred. 
This could be related to the fact that, in the Heisenberg limit $A=1$, the weathervane modes described in figure 1 can rotate about any axis whereas for $A>1$ these excitations select the $z$-axis. Monte Carlo snapshots of the spin configurations at
 very low  temperatures   reveal an exotic phase for which there is  no evidence of long-ranged spatial order of the individual spins\cite{n8}. Both weathervane modes and other localized defects are observed in these snapshots. Figure 2 shows a typical configuration at very low $T$ in the case $A=2$. The lower panel shows the $x-y$ components of the individual spins and the upper panel shows the corresponding $z$ component of the magnetization on each upward traingle. A localized defect is observable which corresponds to a triangle with two spins up and one down and zero chirality. Both the magnetization and heat capacity exhibit critical behaviour at a finite temperature $T_ c$ corresponding to the breaking of the $z$-axis up-down symmetry  and is similar to the $2d$ {\it ferromagnetic} Ising model.

  We employ Monte Carlo methods using a single spin flip heat bath algorithm to  study
 lattices containing N spins with periodic boundary conditions.  
The number of spins is related to the number of unit cells as $N=3L^2$, where $L$ is the number of up-triangles in the
horizontal direction.
  We have calculated various thermodynamic observables such as the
 internal energy, the specific heat, the z-component of the magnetization as well as
 the associated susceptibility and Binder cumulant \cite{n13}. 
 Our numerical data are analyzed
 by using finite size scaling theory and the histogram method \cite{n6b,n15a,n15b} to extract the
 critical exponents of this model. A reweighting method is combined with our
 single spin flip algorithm in order to obtain the observables as continuous functions of
 temperature near $T_c$. We  first measure the specific heat using the
 MC simulations on a discrete temperature grid and this step yields an estimate of the temperature $T_0$
 at which the specific heat is maximum. Using this estimate, the  histogram
 $\Omega_0({E})$  of the number of  spin states with energy $E$ is constructed from MC runs at
 the temperature $T_0$ over a large time interval $\Delta t$. This procedure allows us to obtain
 the average value of any observable Q as a continuous function of temperature $T$
 near $T_0$ as follows
 \begin{equation}
 <Q>=\frac{\sum_{{E}}Q({E}) \Omega_0({E}) e^{-(T^{-1}-T_{0}^{-1})E}}
 	{\sum_{{E}}\Omega_0({E}) e^{-(T^{-1}-T_{0}^{-1})E}}
 \end{equation}
where $Q(E)$ is the microcanonical average of the observable. 
 This method has been used quite succesfully to extract critical exponents of both
 discrete \cite{n15a}and continuous \cite{n6b,n15b,n15c} spin models. 
  
  We have also studied dynamical properties of this model by considering  the 
double-time spin-spin autocorrelation function 
 \begin{equation}
C(t,t_w)=\frac{1}{N}<\sum_i S_{i}^{z}(t_w)S_{i}^{z}(t+t_w)>
\end{equation}
To measure this quantity at a given temperature $T$ we start from a random
configuration at high temperature and rapidly quench to the working temperature $T$. 
We then
wait for a time $t_w$ and measure the autocorrelation function $C(t,t_w)$ for
subsequent times t. The results are averaged  over many random initial
states. In equilibrium one expects $C(t,t_w)$ to be independent of $t_w$
and it is only in equilibrium that one can define a meaningful timescale
associated with relaxation. In the aging regime, $C(t,t_w)$ is waiting 
time dependent \cite{n16a,n16b}. Aging is a  general phenomenon which occurs in a wide variety of off-equilibrium materials,  as 
for example glasses.  The phenomenon has been widely studied in disordered 
systems such as spin glasses \cite{n11}, frustrated systems \cite{n12a} and in 
the phase ordering kinetics of  the Ising ferromagnet \cite{n12c}, and is associated with a slow domain dynamics.
 
\begin{figure}[t]
\centering
\includegraphics[height=60mm,width=80mm]{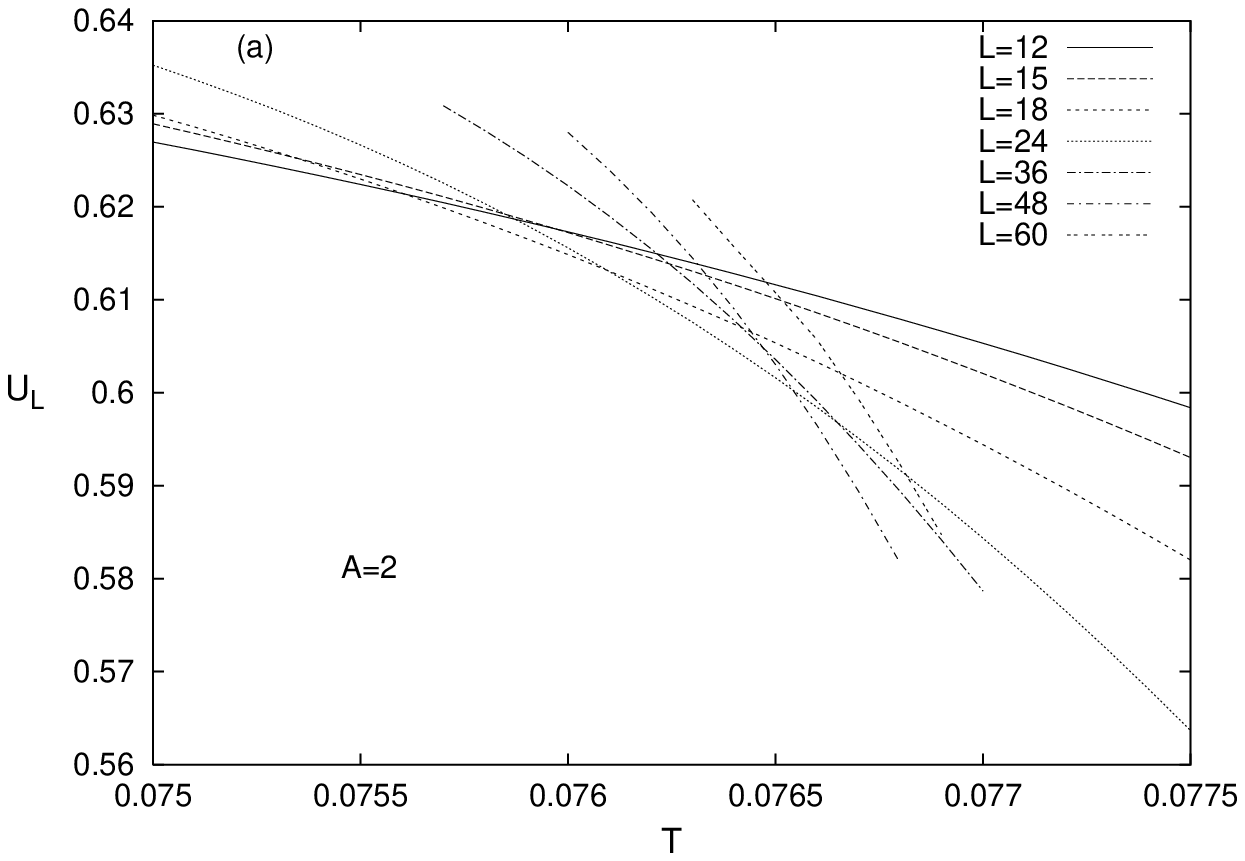}
\includegraphics[height=60mm,width=80mm]{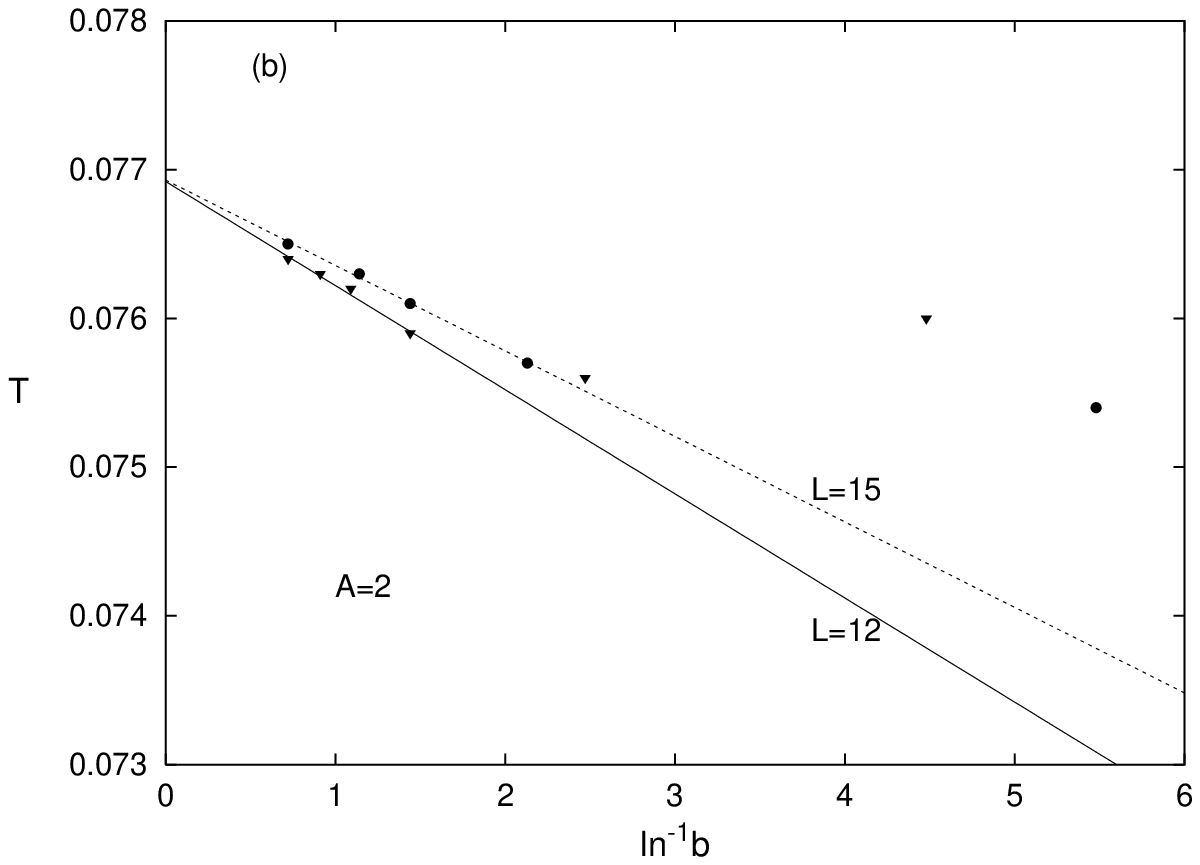}
\caption{ (a) The order-parameter Binder cumulant $U_L$ ($12 \leq L \leq 60)$ plotted vs $T$
obtained by optimized reweighting in the case A=2. (b) Estimates for $T_c$ plotted vs inverse
logarithm of the scale factor $b=L'/L$.}
\end{figure}

\begin{figure*}[t]
\begin{minipage}{80mm}
\center{\includegraphics[
        height=55mm,width=80mm,
        ]{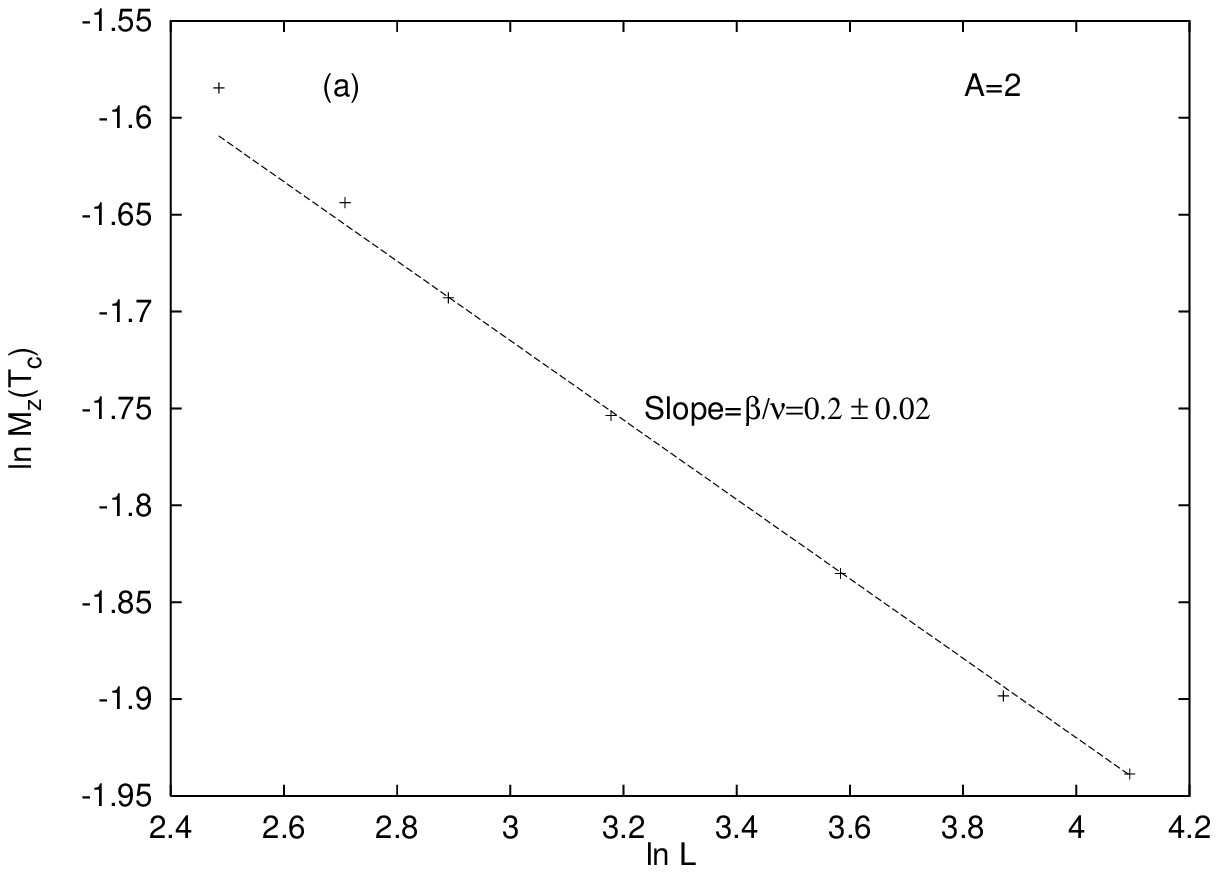}}
\center{\includegraphics[
        height=55mm,width=80mm,
        ]{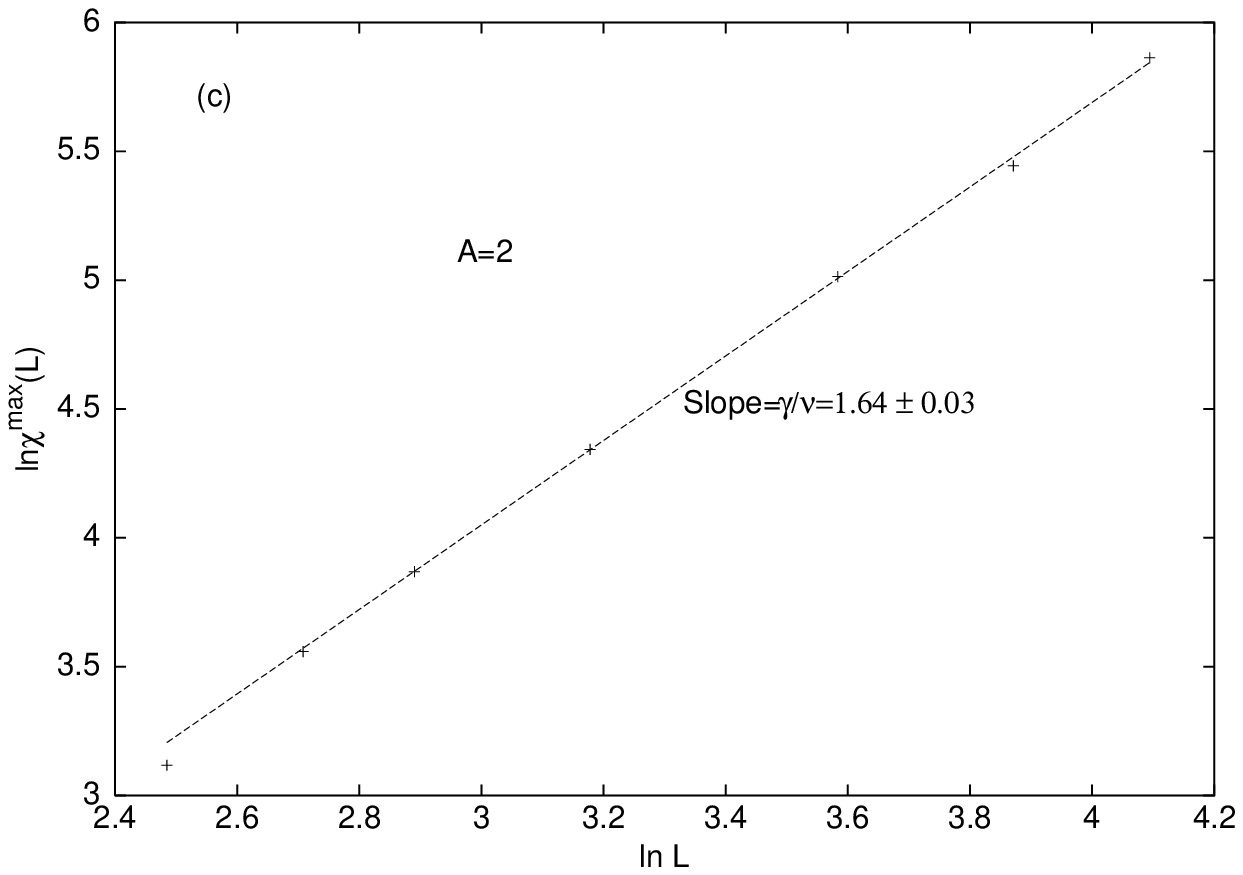}}

\end{minipage}
\hspace{14pt}
\begin{minipage}{80mm}
\center{\includegraphics[
         height=55mm,width=80mm,
        ]{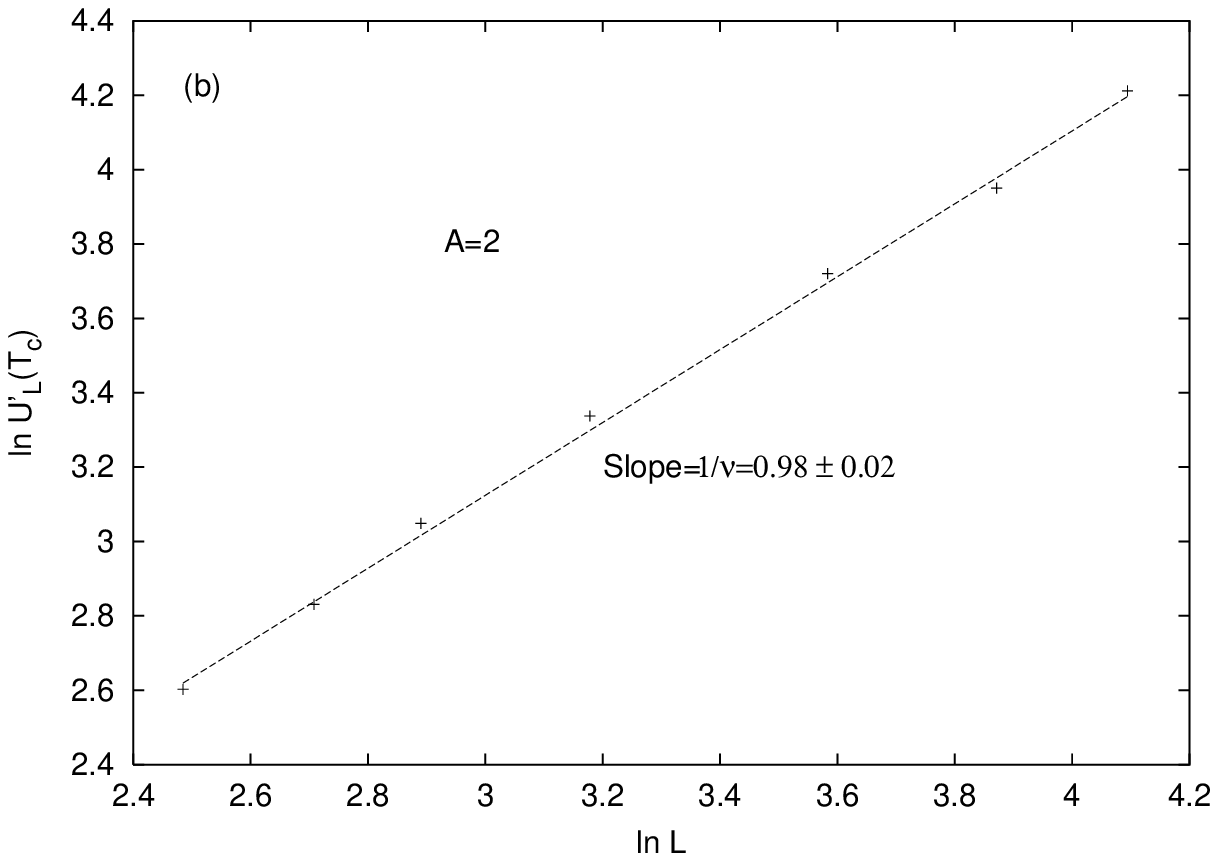}}
\center{\includegraphics[
       height=55mm, width=80mm,
        ]{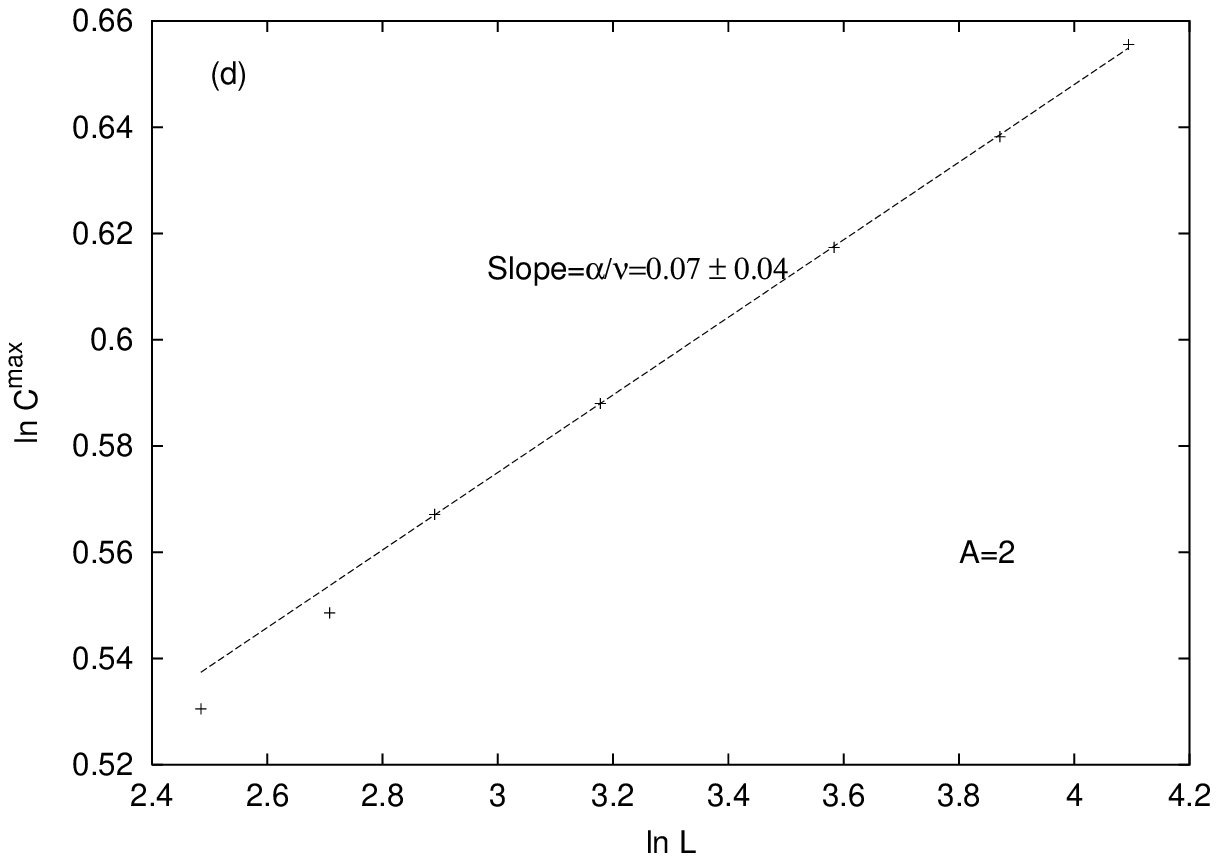}}

\end{minipage}
\caption{Finite size scaling dependence of the critical properties for $A=2$   (a) The order parameter $M_z$ at $T_c$, (b) the
temperature derivative of the Binder cumulant associated with $M_z$ at $T_c$, (c) the
maximum of the susceptibility $\chi^{max}$ and (d) the specific heat maximum
$C^{max}$  plotted as a function of $L$ on a log-log scale.}
 \end{figure*}

\section{Equilibrium Properties}
  We have used our MC method to study lattice sizes  $L=6, 12, 18, 24, 36, 48, 60$ and have used $1-5 \times 10^5$ Monte Carlo steps(MCS) for 
performing our measurements after discarding
the first $5 \times 10^4$ MCS to reach thermal equilibrium. In the reweighting analysis, it is
 important to take $\Delta t$ as large as possible to have good
 statistics. We have used $\Delta t=1.2 \times 10^6$ to $\Delta t=2.6 \times 10^6$
 MCS for small and large lattice sizes respectively. Also, because the energy
 is continuous, we have used both $10 000$ and $30 000$ bins for the histograms in order to check that the size of
 the bins did not affect our numerical results. In the results that follow, the magnitude of the exchange constant $J$
is set equal to unity.

  The critical temperature, $T_c$, can be determined by comparing the reduced Binder cumulant of the magnetization,
$U_L = 1 -{<M_z^4>}/3{<M_z^2>^2}$, for
  lattices of  size $L$ with lattices of  size $L'=bL$ as shown in figure 3(a) for the case of the anisotropy parameter $A=2$.
  In the limit of large system sizes, the cumulants
should cross at the critical temperature\cite{n16c,n16d} and have a  common value  $U_L=U^*$  . 
However, due to finite size effects, it is necessary to
  extrapolate the crossing points to the limit $b \rightarrow \infty$ \cite{n16d}.
  Our results using the Binder cumulant crossing method \cite{n13} to estimate the critical
  temperature are presented in figure 3(b). The points represent the temperatures at
  which the order parameter cumulant for $L'$ crosses the cumulant for
  $L=12$ or $L=15$. There is considerable scatter in the data and care must be taken
  to use only results with $L'$ sufficiently large to be in the asymptotic region
  where a linear extrapolation is justified ($\frac{1}{\ln b} \leq 2.2$). Using
  this method, the critical temperature is estimated to be $T_{c}=0.077 \pm
  0.001$. This
  value is slightly lower than the one obtained by KM  from phenomenological
  renormalization of the magnetization but lies  within their error bars.
  No hysteresis is observed in the order parameter nor in the
  energy near the critical region. In addition, no double peak structure was
  found in the energy histograms and  the Binder energy cumulant 
  evaluated  at $T_c$ yielded  the result  $U^*=0.666665(7)$ for large $L$,
  consistent with the value $\frac{2}{3}$ expected for a continuous transition
  \cite{n16e}.

  Finite-size scaling results for the order parameter $M_z$,
  the first temperature derivative of its Binder cumulant, the susceptibility 
  $\chi=\frac{N}{T}(<M_{z}^{2}>-<M_{z}>^2)$ and the specific heat are
  shown in figures  4 (a-d) respectively on a log-log scale. According to the standard theory of 
  finite size scaling,
  the equilibrium magnetization $M_z$ should obey the relation $M_{z} \sim
  L^{-\beta/\nu}$ for sufficiently large $L$. Figure 4(a) shows our results of a finite
  size scaling analysis for the order parameter $M_z$.  Excluding the smallest two
  lattice sizes $L=12$ and $15$ from the fitting procedure, we obtained the value of
  the exponent  ratio $\beta/\nu=0.20 \pm 0.02$ which is significantly larger than the 2d Ising
  value $\beta/\nu=1/8$ .

\begin{table}[t]

 \caption{Results for the static critical temperature $T_c$  and the 
 exponents $\gamma/\nu$,  $\beta/\nu$ and  $\nu$  for various values of
 the anisotropy $A$. }
  \begin{ruledtabular}
  \begin{tabular}{|c|c|c|c|c|}
\hline
A &$T_c$&$\gamma/\nu$&$\beta/\nu$&$1/\nu$ \\
\hline 
1.1& $0.036\pm 0.006$ &$1.44\pm 0.07$ &$0.30\pm 0.05$ &$0.97 \pm
0.06$\\
\hline 
1.5& $0.067\pm 0.001$ &$1.61\pm 0.05$ &$0.22\pm 0.04$ &$1.03 \pm
0.04$\\
\hline 
2& $0.077\pm 0.001$ &$1.64\pm 0.03$ &$0.20\pm 0.02$ &$0.98 \pm
0.02$\\
\hline 
3& $0.076\pm 0.001$ &$1.66\pm 0.05$ &$0.18\pm 0.03$ &$0.99 \pm
0.03$\\
\hline 
5& $0.064\pm 0.002$ &$1.67\pm 0.04$ &$0.17\pm 0.03$ &$1.01 \pm
0.03$\\
\hline
8&$ 0.052 \pm 0.003$& $1.67\pm 0.04 $& $0.16 \pm 0.01$& $1.09 \pm 0.06$  \\
\hline 
14& $0.037\pm 0.001$ &$1.70\pm 0.05$ &$0.14\pm 0.01$ &$1.02 \pm
0.04$\\
\hline 
20& $0.030\pm 0.001$ &$1.73\pm 0.03$ &$0.13\pm 0.01$ &$1.03 \pm
0.04$\\
\hline
30&$ 0.022 \pm 0.002$ &$1.72 \pm 0.04$&$ 0.13 \pm 0.02$& $1.02 \pm 0.04$ \\
\hline
 \end{tabular}
 \end{ruledtabular}
 \end{table}

   The behavior of the reduced Binder cumulant $U_L$ at the critical
   point
  can be used to find the value of the critical exponent $\nu$. Finite size
  scaling theory predicts   at $T_c$ that $U_{L}=U_0(tL^{1/\nu})$ with $t=\mid
  1-T/T_{c}\mid$ and the temperature derivative of $U_L$ at $T_c$ should obey the relation
  $U'_{L}(T_c)=L^{1/\nu}U'_{0}(0)$. In figure 4(b) we show that this prediction is
  obeyed quite well. The value of the static exponent $\nu$ obtained using a least-squares
  fit is $1/{\nu}=0.98 \pm 0.02$, which is remarkably close to the two dimensional 
  Ising value $\nu=1$.
  The magnetic susceptibility $\chi$  has 
  the scaling form $\chi \sim L^{\gamma/\nu}$ and figure 4(c) shows
a least squares fit to our results using this form and
  we find $\gamma/\nu=1.64 \pm 0.03$ which is smaller 
  than  the 2d Ising value $\gamma/\nu=7/4$ and also the value obtained by KM.
  The specific heat was also calculated 
  but it
  is much more difficult to analyze because of the small number of points used 
  and the scatter in the data was too large to extract a reliable
  estimate for $\alpha/\nu$ .
  $C_{max}$ should scale in the critical region as $C_{L}^{max}\simeq
  C_0+aL^{\alpha/\nu}$, with $C_0$ representing the regular part. From our 
  fitting procedure, $C_0=0$ yields the best straight line for large  sizes with  slope 
  $\alpha/\nu=0.07 \pm 0.04$ . 


\begin{figure}[htp]
\centering
{\includegraphics[height=70mm,width=85mm]{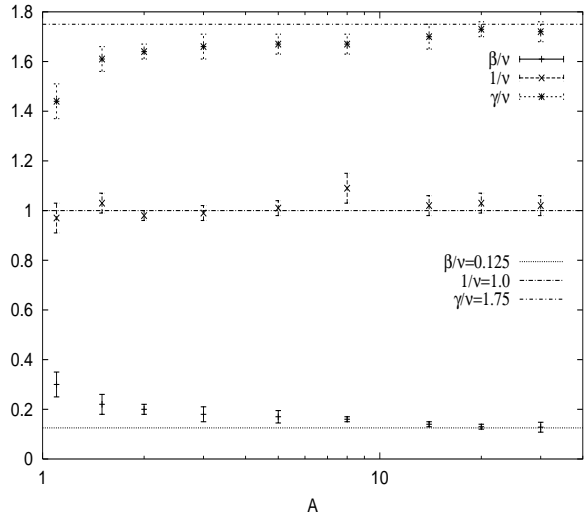}}
\caption{Variation of the critical exponents with the anisotropy $A$ obtained from the equilibrium properties. The symbols
represent the measured values and the lines indicate the values expected for the $2d$ Ising model.}
\end{figure}

  The same analysis has been carried out for other values of the
  anisotropy  $A$ and the results are summarized in Table I. The critical temperature  increases  from zero for small
  deviations  of $A$ from unity, attaining a maximum  near $A\sim 2$ and decreasing to very low temperatures for
  large $A$ in agreement with KM.  In the limit $A \rightarrow \infty$, the transition temperature approaches zero as in the case of the spin $1/2$ antiferromagnetic Ising model. It is not clear whether or not there is an ordered ground state in the limit $A \rightarrow \infty$. The critical exponents are plotted as a function of the anisotropy parameter $A$ in figure 5.
 The values of the critical exponents  $\beta/\nu$ and $\gamma/\nu$ appear to depend on the value of $A$  and approach
  the usual 2d  ferromagnetic Ising values  
  as $A$ becomes very large.  Both universality and weak universality\cite{n28} are violated in this system. This non-universal behaviour of
  Ising-like exponents 
  has also been reported for a two dimensional system of $XY$ spins
  interacting  via both ferromagnetic and antiferromagnetic bonds in the presence of
  an applied magnetic field which reduces the  symmetry $O(2)$ in spin
  space to $Z_2$ \cite{n17}.

\begin{figure}[t]
\centering
\includegraphics[height=60mm,width=85mm]{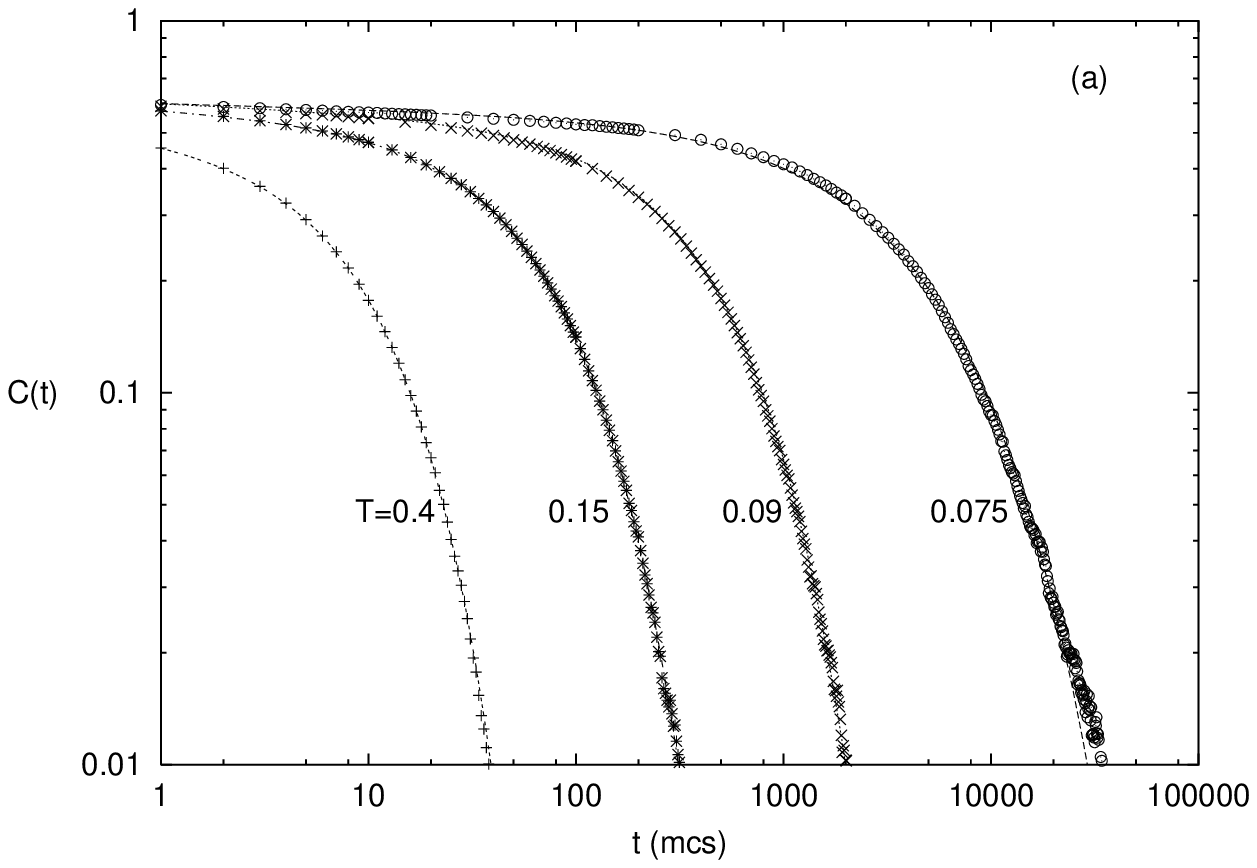}
\includegraphics[height=60mm,width=85mm]{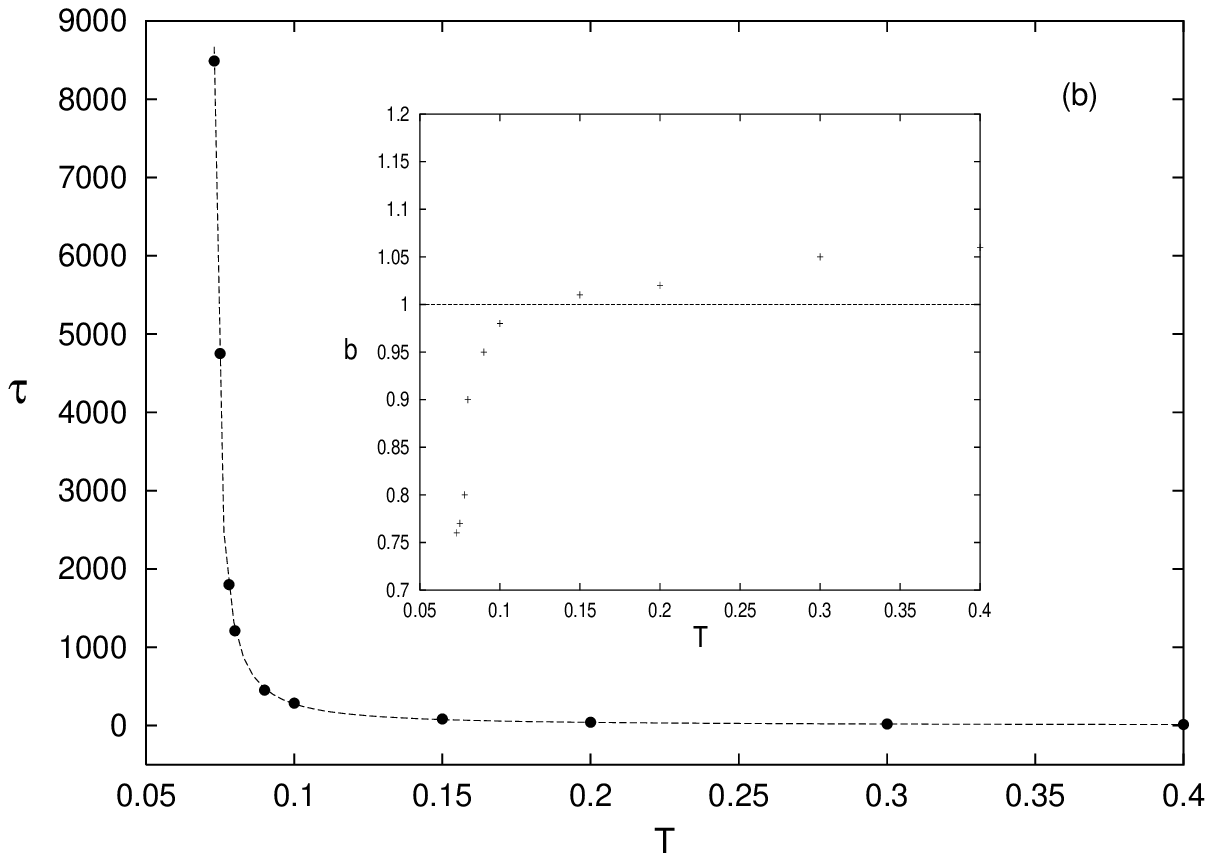}
\caption{(a) The spin-spin autocorrelation function for various temperatures,
$T=0.4 (+)$, $T=0.15 (\ast)$, $T=0.09 (\times)$,$T=0.075 (\circ)$ for $A=2$.
The lines are fits to form given in (5) (b) The associated relaxation time as a function
the temperature together with a fit to the power-law form (6). The inset
shows the temperature dependence of the exponent $b(T)$.}
\end{figure}

  In order to understand  the nature of this exotic structure in more detail, we
  have also looked at the spin-spin autocorrelation function using equilibrium
  dynamics at high temperatures where $C(t,t_w)$ becomes independent of $t_w$
  for $t_w>10^4$. As the temperature is reduced, the relaxation of the spins becomes slower and
deviates from a simple exponential form. 
We fit our data with the following function\cite{n11}, 
  \begin{equation}
  f_0(t)=\frac{a}{t^{x(T)}} exp(-(t/\tau)^{b(T)})
  \end{equation} 
  where $a, b(T), \tau(T), x(T)$ are all fitting parameters. As shown in figure 6(a) for $A=2$, our data are
  fairly well described by this functional form and we are able to
  extract the temperature dependent relaxation time  $\tau(T)$  and the exponent
  $b(T)$. The behavior of these parameters is depicted in figure 6(b). The relaxation  time $\tau$
  appears to increase sharply as $T$ is reduced and can be fit quite accurately 
  by a power law divergence of the form
  \begin{equation}
  \tau \sim (T-T_g)^{-\phi}
  \end{equation}
  The glass temperature, $T_{g}$, identified from the power law divergence of the
  relaxation time is slightly lower than the critical temperature $T_{c}$ obtained from the
  equilibrium  measurements for values of  $A$ which are not too large.  For $A=2$,  it seems that there is a
  decoupling between the local spin  degrees of freedom and the net magnetization on each triangle.  The composite spin
variable orders at $T_c$ and the local spin variables enter a glassy phase at $T_g < T_c$. At larger values of $A$ 
  the glass temperature $T_g$ approaches the critical temperature $T_c$. 
  As  will be shown in the next section, $T_g$ signifies the onset of aging 
  phenomena and it is impossible to define a timescale
  below this temperature since the system is in a frozen glassy  state.

\begin{table}[t]

 \caption{Results  obtained from high
 temperature equilibrium dynamics of the spin-spin correlation function $C(t,t_w)$: $T_c$ is the transition temperature
obtained from the equilibrium properties, $T^*$ is the
 temperature  at which $C(t,t_w)$ first has a non-exponential behaviour; $T_g$ is the 
 critical temperature obtained from (6) where the relaxation time diverges; $b_1$ is the
 lowest value of the  exponent $b(T)$ in (5) near $T_g$ and $\phi$ is the relaxation time
 exponent in (6).}
  \begin{ruledtabular}
  \begin{tabular}{|c|c|c|c|c|c|}
\hline
A &$T_c$&$T_g$&$T^*$&$b_1$&$\phi$ \\
\hline 
1.1& $0.036\pm 0.006$ &$0.032\pm 0.001$&$0.050$&0.76&$1.5$\\
\hline 
2& $0.077\pm 0.001$ &$0.071\pm 0.002$&$0.100$&0.75&$1.3$\\
\hline
8&$ 0.052 \pm 0.003$& $0.05\pm 0.001$ & $0.055$ & $0.85$ & $0.74$ \\
\hline
30&$ 0.022 \pm 0.002$ & $0.022\pm 0.001$&$0.026$&0.86&$ 0.55$ \\
\hline
 \end{tabular}
 \end{ruledtabular}
 \end{table}

  The exponent $b(T)$ is plotted in the inset of figure 6(b). It is temperature
  dependent and lies in the range $b_1<b<1$ for $T_g<T<T^*$ where $b_1=0.75,
   T_g=0.071$ and $T^*=0.1$ for $A=2$ . The
  non-exponential behavior sets in at temperatures below $T^*$ . 
  The  parameter $x(T)$ which characterizes the short time behaviour lies in the range $0 < x(T) < 0.1$ and decreases
  with temperature. A summary of our  relaxation results  for a few values of $A$ are given in Table II.

The relaxation time exponent $\phi$ is also nonuniversal  and decreases in value for larger values of $A$.   
   Both non-exponential relaxation  and a  diverging relaxation time are
  features of glasses \cite{n11} and this behaviour has previously been seen in frustrated systems
  without disorder \cite{n12a}.



\begin{figure*}[htp]
\begin{minipage}{85mm}
\centering
\includegraphics[
        height=70mm,width=85mm,
        ]{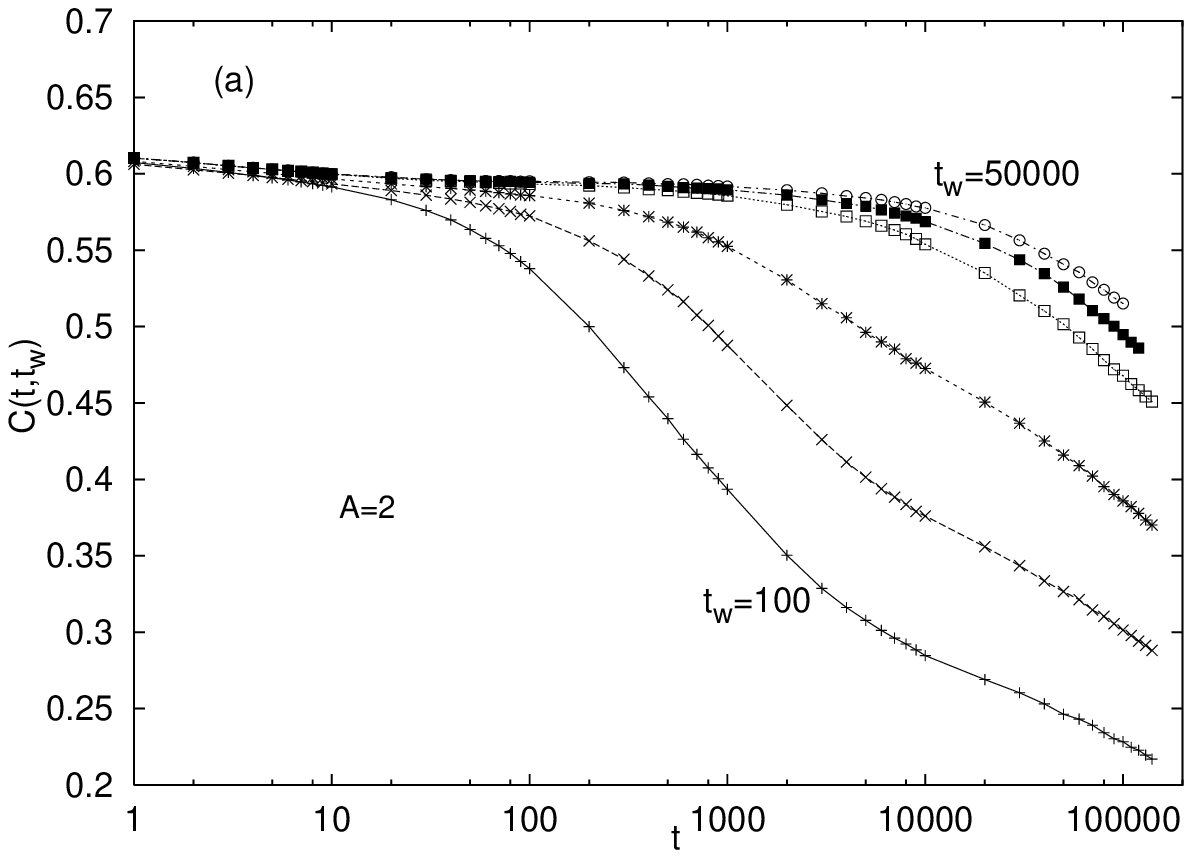}

\end{minipage}
\hspace{14pt}
\begin{minipage}{85mm}
\center{\includegraphics[
         height=70mm,width=85mm,
        ]{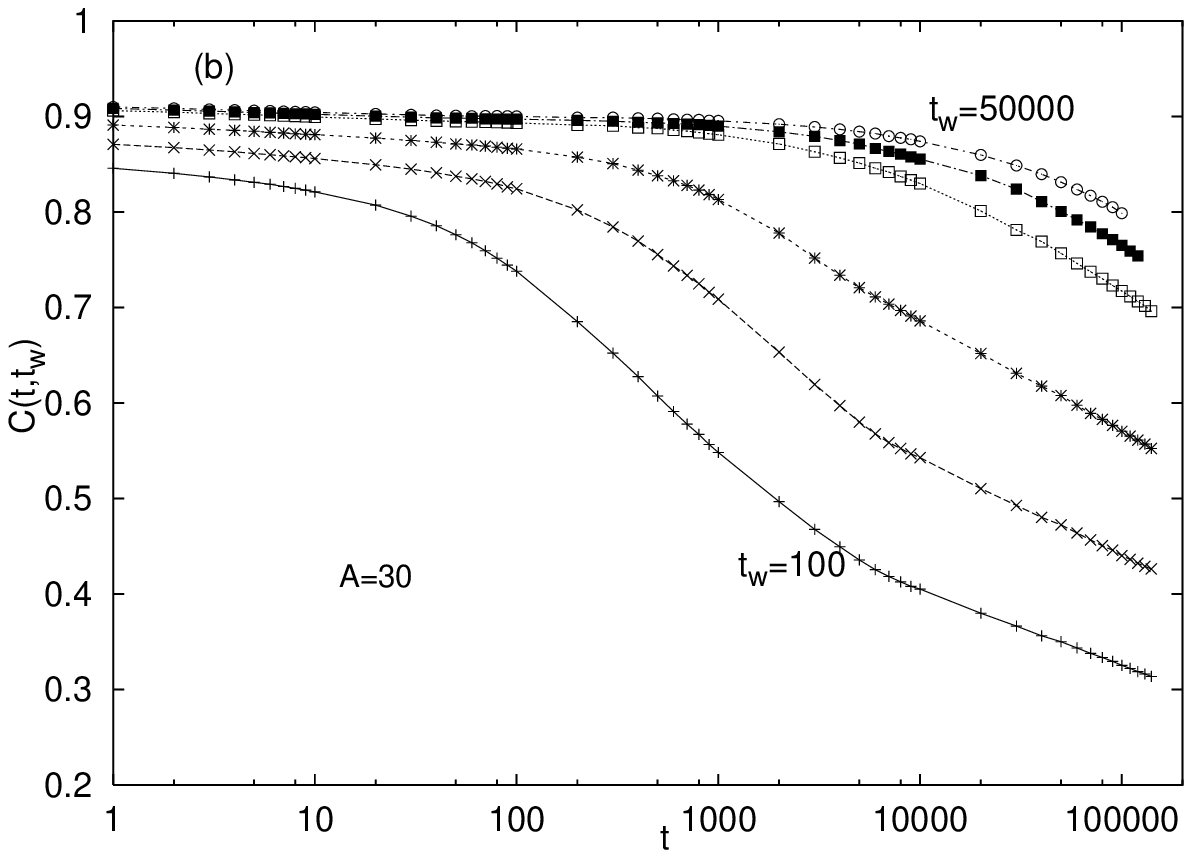}}

\end{minipage}
\caption{Autocorrelation function $C(t,t_w)$ vs the observation time t for
different waiting times $t_w$ from bottom to top $t_w=100, 500, 2000, 
10000, 25000,50000$ at (a) $T=0.04$ for $A=2$ and (b) $T=0.01$ for $A=30$.
}

 \end{figure*}


 \section{Off-equilibrium dynamics}
   In order to further study the dynamics of this model, we have carried out some
   numerical experiments focused on revealing the presence of slow dynamics in
   conjunction with history-dependent phenomena which is generally referred to as  'aging'. These features are
   most easily found  in simulations of the two-time
   autocorrelation function $C(t,t_w)$ which  shows an explicit
   dependence on both times $t,t_w$ over a wide range of time scales. Aging can be
   observed in real systems through different experiments. A typical example is
   the zero-field cooling experiments in which the sample is cooled in zero
   field to a subcritical temperature at time $t=0$. After a waiting time $t_w$ a
   small magnetic field is applied and subsquently the time evolution of the
   magnetization is recorded. It is often observed that  the relaxation becomes slower as  the waiting
   time $t_w$ is increased.

   We have measured the behavior of $C(t,t_w)$ as a function of the observation time $t$,
   for different values of $t_w, A$ and $T$. We have used  $1.5 \times 10^5$
   MCS  with a lattice size $L=36$ and averaged the results over approximately100 different trials. 
      At high temperatures $T>T_g$, we found that the system does not exhibit
   aging since for any value of $A$  the autocorrelation function$C(t,t_w)$ is homogeneous in time
   and independent of $t_w$.

   In figures 7(a-b) , the behaviour of the autocorrelation function clearly confirms 
   the presence of aging in this model for all values of $A>1$  at very low
   temperatures $T<T_g$. For large waiting times and $t<<t_w$, the correlations are  independent of the waiting time $t_w$.
   However, for $t>t_w$, the curves show an explicit 
   dependence on both times indicating
   that equilibrium has not been attained within the time of the simulation  and the correlation falls to 0 for $t \rightarrow \infty$. This
   scenario has been called weak ergodicity breaking \cite{n19a,n19b}. The fluctuation
   dissipation theorem holds for short times but is violated at longer times.

\begin{figure}[b]
\centering
\includegraphics[height=70mm,width=85mm]{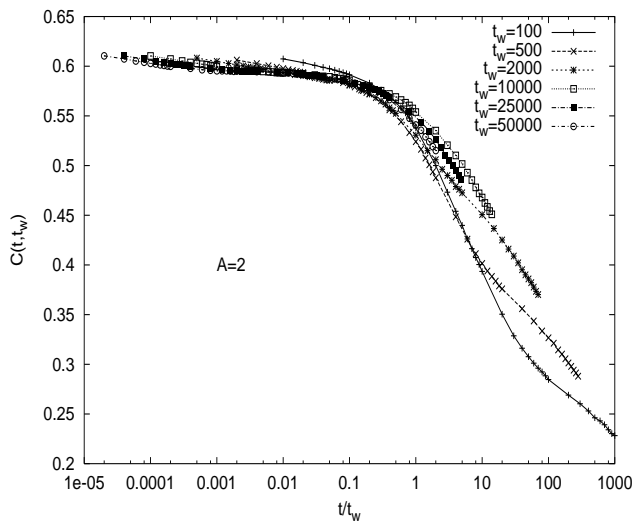}
\caption{Autocorrelation function $C(t,t_w)$ vs $t/t_w$ at $T=0.04$, for $A=2$. 
}
\end{figure}

\begin{figure}[t]
\centering
\includegraphics[
        height=70mm,width=85mm,
        ]{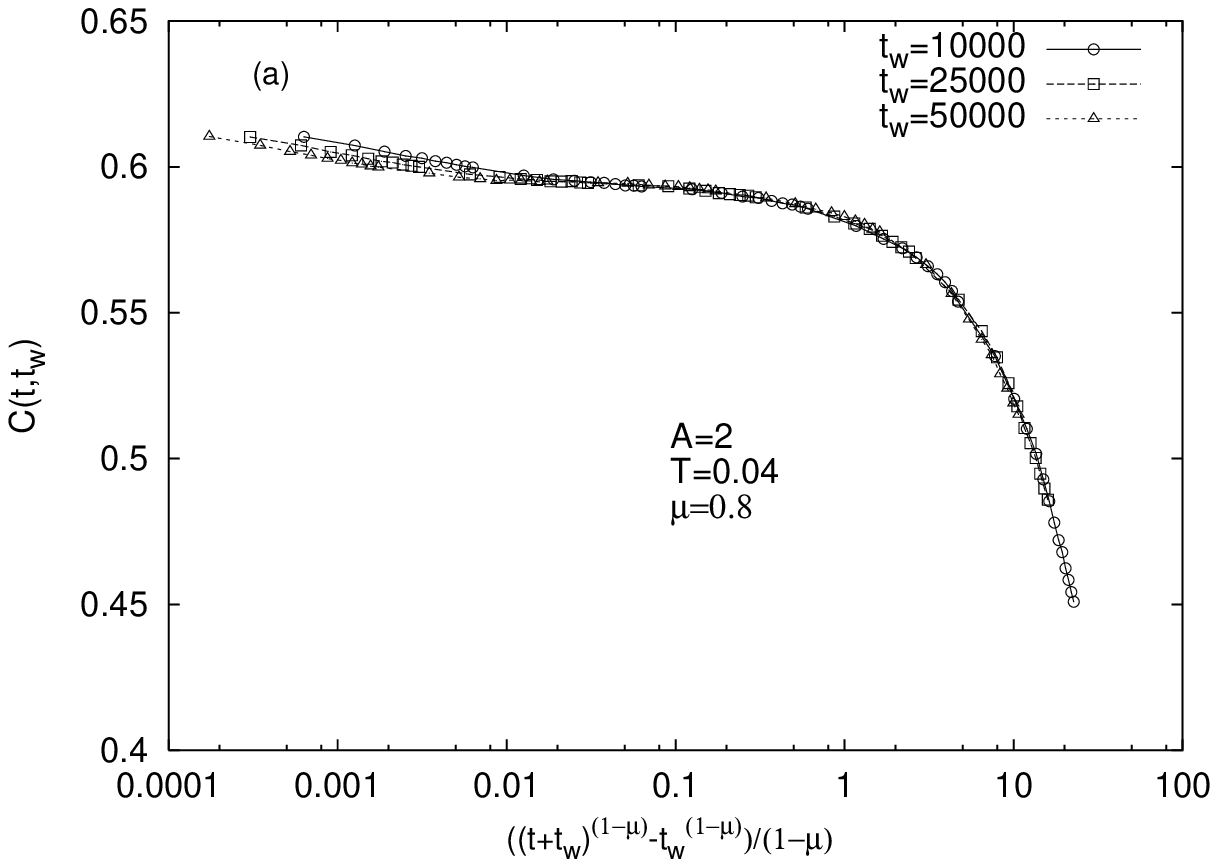}

\center{\includegraphics[
         height=70mm,width=85mm,
        ]{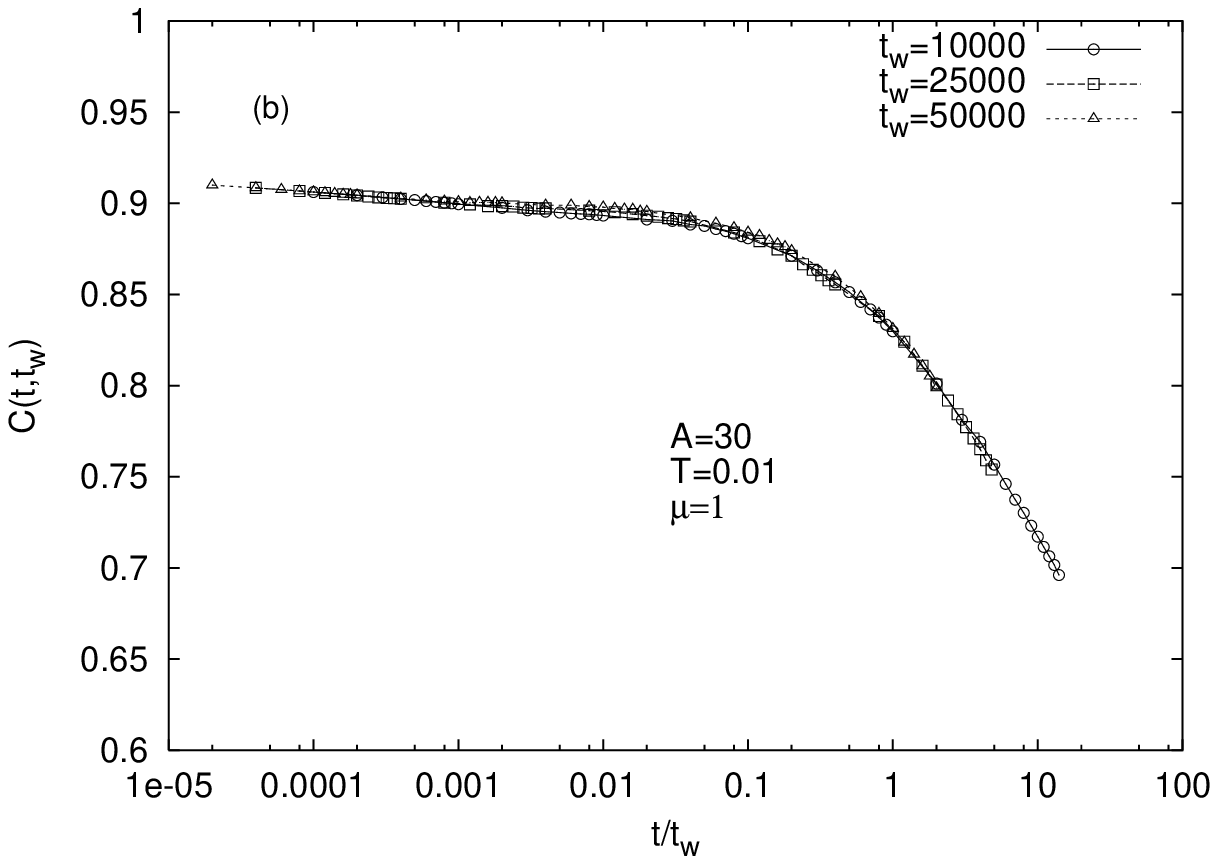}}

\caption{Data collapse of the curves shown in figures 7(a,b) at large waiting times
for (a) $T=0.04$ and $A=2$ as a function of the time reduced variable 
$[(t+t_w)^{1-\mu}-t_{w}^{1-\mu}]/(1-\mu)$ with $\mu=0.8$ and (b) $T=0.01$ and $A=30$
as a function of $t/t_w$. 
}

 \end{figure}


   We have attempted to find an appropriate scaling law for the
   aging curves. A knowledge of the scaling form  could give some  insight into the nature of
   the underlying dynamical process, even if there is no theoretical basis for
   determining the scaling functions. The simplest scenario 
   is naive aging of the form 
   \begin{equation}
   C(t,t_w)=f(\frac{t}{t_w})
   \end{equation}
   in the region where both $t$ and $t_w$ are large.

	In figure 8 we show  $C(t,t_w)$ as a function of $t/t_w$ to see if this naive form of scaling holds. 
 Except for the largest value of $t_w=50000$, we observe a departure from naive scaling with the
 function $C(t,t_w)$   increasing when $t >t_w$ and decreasing for $t < t_w$  
 at fixed values of $t/t_w$ as $t_w$ increases.
 A 'superaging'
 behaviour as observed in mean field spin glasses or the Sherrington Kirkpatrick (SK) model \cite{n20c} 
 could be expected for smaller waiting times but  
    the fact that the curve for the largest 
 waiting time,  $t_w=50000$, lies below the next smaller value, $t_w=25000$, could be
 explained as follows:  the system is in a 'subaging' scaling region where the relaxation of
 older systems becomes faster when plotted versus $t/t_w$ although, when plotted
 versus $t$, the older the system appears to exhibit slower  relaxation.
Indeed,
  as seen in figure 9(a), a good collapse in the asymptotic region of the 
  largest
 waiting times is obtained by using a variable  used in glassy
 polymers \cite{n21} and recently in a topological spin glass \cite{n10,n19b}, namely
 $[(t+t_w)^{1-\mu}-t_{w}^{1-\mu}]/(1-\mu)$
 where  in our case the value $\mu=0.8 <1$ is used. This subaging effect has not only been observed in glassy
 polymers \cite{n21} but also in 2d  site-disordered spin glasses\cite{n22}. 

  For larger values of the anisotropy, our  analysis of the largest waiting
  times shows that simple scaling holds, compatible with the full aging scenario. 
Figure 9(b) shows the results for $A=30$.  This simple scaling has also been observed in the 3d Edwards-Anderson spin glass
\cite{n19a,n19b,n23}. 


\begin{figure}[b]
\centering
{\includegraphics[height=70mm,width=85mm]{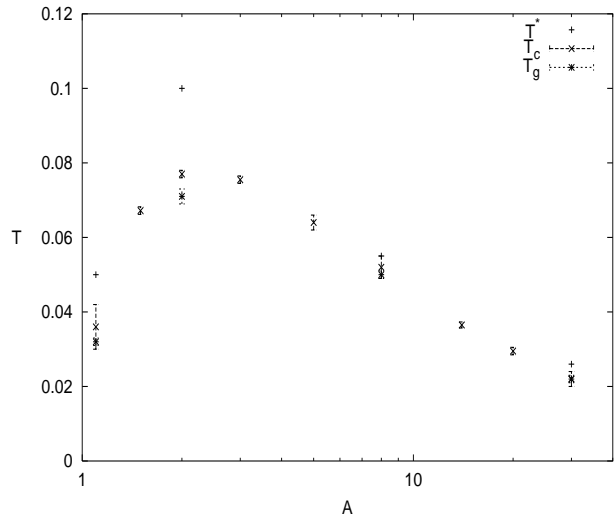}}
\caption{The temperatures $T^*, T_c$ and $T_g$ plotted  as a function of the anisotropy $A$ using the values listed in Tables I and II.}
\end{figure}

\section{Summary}
	In this work we have performed a  numerical study of the two dimensional easy-axis Heisenberg 
antiferromagnet on the Kagome lattice by computing its static and dynamic properties. From the static properties,
we have extracted the critical temperature and the critical exponents associated with the
magnetization, the susceptibility and the correlation length respectively. Our result for the critical
temperature obtained from the Binder cumulant method is in agreement with that obtained by KM \cite{n8}
within statistical errors. On the other hand, our results for the critical exponents indicate that this
system has a nonuniversal phase transition. Namely, the values of the exponents $\beta$ and $\gamma$ associated with the order parameter and the
susceptibility   vary with the magnitude of the easy-axis anisotropy $A$. However, $\alpha$ and $\nu$ remain unchanged
and correspond within errors to the 2d  Ising values and thus weak universality is also violated. Although the magnetization indicates a finite $T_c$, Monte Carlo
snapshots of the individual spins below this temperature do not indicate any long ranged spatial order. Rather, the individual spins
appear to be in a frozen state similar to a glass.

  	We have  studied the two-time spin-spin correlation function, $C(t,t_w)$, at high and low temperatures
 respectively. We have found that the high temperature equilibrium correlation function is described very well  by the
 function $a t^{-x} exp(-(t/\tau)^b)$ suggested by Ogielski \cite{n11} over the entire time and
 temperature range. Non-exponential relaxation sets in at a temperature $T^* > T_c$. The relaxation time, $\tau$, increases according  to a power law and diverges at a
 temperature $T_g < T_c$ where a transition to a glassy phase is located.  Figure 10 shows our results obtained for the temperatures
$T_c, T_g$ and $T^*$ obtained from the statics and dynamics plotted as a function of $A$.  KM also identified a broad maximum in the specific heat for $A > 2$ at a lower temperature than those in figure 10 which they attribute to a local degree of freedeom since the peak does not exhibit any size dependence. This local degree of freedom could be the weathervane mode or the defect observed in figure 2.  

Maegawa et. al.\cite{n27} have reported
an observation of successive phase transitions in the Kagome systems $RFe_3(OH)_6(SO_4)_2 [R=NH_4, Na, K]$. Susceptibility cusps are observed at two closely spaced temperatures which are about $10\% $ of the corresponding Curie Weiss temperatures. These transitions
could be explained in terms of the ordering of the magnetization on each triangle at the upper temperature $T_c$ followed by a spin freezing
of the local spins at the lower temperature $T_g$.

Below $T_g$, 
  we have found clear
 evidence for the presence of aging effects in the autocorrelation function from
 off-equilibrium dynamics. The spin-spin autocorrelation function depends on both times and the dynamics
 becomes slower for larger waiting times. An analysis of the autocorrelation functions from scaling forms used in 
 polymer glasses and
  spin glasses has shown different behaviour. Namely, a sub-aging behavior at low values of $A$
  is seen where the relaxation time of the system grows more slowly than the waiting time $t_w$ as observed in  2d spin glasses
  \cite{n22}, polymer glasses \cite{n21} and structural glass
  \cite{n24}, whereas for  large values of $A$,
  a full aging behavior describes the data well, where the relaxation time of the system scales with its age
  $t_w$ as observed in 3d spin glasses \cite{n19a,n23}.
We are currently extending our investigations of this system to include the effects of a small applied magnetic field
on the aging behaviour. This will enable us to check whether the fluctuation dissipation theorem is
   violated  and to study the long-term memory of this model.

\begin{acknowledgments}

This work was supported by the Natural Sciences and Research Council of Canada 
and the High Performance Computing facility at the University of Manitoba. We would also
like to thank Walter Stephan and Chris Henley for many useful discussions.
\end{acknowledgments}

%
%
\bibliography{smaine}

\end{document}